\documentclass[11pt,onecolumn]{article}
\usepackage{amssymb,amsfonts,amsmath}
\usepackage{algorithmic}
\usepackage{array}
\usepackage{mdwmath}
\usepackage{mdwtab}
\usepackage{fixltx2e}
\usepackage{amsthm}
\usepackage{setspace}
\usepackage{latexsym}
\usepackage{mathtools}
\usepackage{graphics}
\usepackage{graphicx}
\usepackage{epsfig}
\usepackage{color}
\usepackage{authblk}
\usepackage{tocbibind}
\usepackage{lineno}
\usepackage{booktabs}

\emergencystretch=\maxdimen
\usepackage[linesnumbered,ruled,vlined]{algorithm2e}

\usepackage[margin=0.7in]{geometry}

\begin{document}
\title{\textbf{Discovering dynamic functional networks in the human neonatal brain with electric source imaging}}
\author[a,b]{Steve Mehrkanoon \thanks{Corresponding author: Steve Mehrkanoon, Building 71/918 RBWH Herston, Brisbane, QLD 4029, Australia. Email: steve.mehrkanoon@unimelb.edu.au and/or stmehrkanoon@gmail.com. The author declares no conflict of interest.}}
\affil[a]{Centre for Clinical Research, The University of Queensland, QLD, Australia}
\affil[b]{Royal Brisbane and Women's Hospital, Metro North Hospital and Health Service, QLD, Australia}
\date{}
\maketitle
Running title: \textbf{Emergent brain network dynamics in the neonates}
\vspace{5mm}
\newpage
\begin{abstract}
When the human brain manifests the birth of organised communication among local and large-scale neuronal populations activity remains undescribed. We report, in resting-state EEG source-estimates of 100 infants at term age, the existence of macro-scale dynamic functional connectivity, which have rich topological organisations, distinct spectral fingerprints and scale-invariance temporal dynamics. These functional networks encompass the default mode, primary sensory-limbic system, thalamo-frontal, thalamo-sensorimotor and visual-limbic system confined in the delta and low-alpha frequency intervals (1-8 Hz). The temporal dynamics of these networks not only are nested within much slower timescale (< 0.1 Hz) but also correlated in a hierarchical leading-following organisation. We show that the anatomically constrained richly organised spatial topologies, spectral contents and temporal fluctuations of resting-state networks reflect an established intrinsic dynamic functional connectome in the human brain at term age. The graph theoretical analysis of the spatial architectures of the networks revealed small-world topology and distinct rich-club organisations of interconnected cortical hubs that exhibit rich synchronous dynamics at multiple timescales. The approach opens new avenues to advance our understanding about the early configuration organisation of dynamic networks in the human brain and offers a novel monitoring platform to investigate functional brain network development in sick preterm infants.
\end{abstract}
\newpage
\section*{Introduction}
\noindent In humans, the development of neuronal systems in late gestation determines organisation of specialised neural functions that underpin large-scale functional organisation of the brain \cite{TauGregory2009,  GRAYSON201715}. As a measure of large-scale functional organisation of resting-state (RS) brain activity, functional connectivity MRI (fcMRI) and electroencephalography (EEG) have revealed richly structured RS networks (RSNs) \cite{Biswal1995537, Mehrkanoon2014BC, Mehrkanoon2014338}, which fluctuate over time and form \textit{dynamic} functional connectome (dFC) in adults \cite{Mehrkanoon2014338, Allen2014663, Boersma21030, Mehrkanoon2016252}. The EEG in infants born between $\sim$ 30-42 wk gestation has shown the emergence of low-frequency neuronal population activity \cite{FranssonPeter102013638, CornelissenLaura2015}, which may exhibit oscillatory activity of large-scale cortical correlation structure measured by EEG-\textit{scalp} FC \cite{FranssonPeter102013638, CornelissenLaura2015, MEIJER2014780, Tokariev20164540, ArichiT2017}. The neuronal oscillations, with their specific frequencies, open the gates of communications between cortical regions to establish patterns of local and large-scale functional brain network interactions \cite{Hipp2012}, which underpin frequency-specific dFC \cite{Hipp2012, Mehrkanoon2014338}. Thus, FC measurements based on the time and frequency components of interacting neuronal population oscillations may provide new insights into the evolutionary macroscopic neuronal circuits and their interaction patterns in the infant brain. Despite numerous RS-fcMRI and EEG studies on the \textit{infant} brain, there is no systematic description on the spatial organisation, spectral fingerprints and temporal dynamics of local and large-scale neuronal populations activity in the infant brain. This may have been limited to the evidence of the infant brain studies: The infant EEG-scalp FC does not reflect the anatomical information of cortical networks \cite{MEIJER2014780, OmidvarniaAmir2014, Schumacher20152261, KaminskaAdoibhx206}, an incomplete anatomical localisation of RSNs \cite{Tokariev20164540, Roche-Labarbe2008, Routier20172345, Mohammadi-Nejad2018}, and the absence of spatial topologies, spectral components and temporal dynamics of RSNs at term age \cite{FranssonPeter102013638, Tokariev20164540, Routier20172345}. To characterise the functional organisation of macroscopic neuronal population activity in infants at term age, we infer anatomically constrained spatial topologies, spectral fingerprints and temporal dynamics of RS-FCs that overcome current limitations in infant EEG-scalp topography.
\section*{Materials and Methods}
\subsection*{Participants}
The Human Research Ethics Committee of the University of Queensland and Royal Brisbane Women's Hospital approved the protocol. All participants gave their informed consent according to the National Health and Medical Research Council guidelines. One-hundred healthy infants participated in this study; 84 preterm-born infants at term-equivalent age (40-42 weeks gestational age), and 16 term-born infants. Informed parental consent was obtained for all participants. Inclusion criteria were preterm birth $<$31 weeks gestational age \cite{FINNIGAN2018143}.
\subsection*{High-density EEG data acquisition}
Surface EEG was acquired during a single 40-min session using a 64-channel neonatal EEG-cap (30-32 cm head circumference) arranged according to the international 10-20 system, EGI Hydrocell GSN 130 Geodesic Sensor Nets, and an EGI Net-Amps-300 amplifier (EGI, Philips, USA) at sampling rate of $f_s=$250 Hz \cite{FINNIGAN2018143}. All EEG data were referenced to Cz. The EEG signals were band-pass filtered (0.5$-$30 Hz) and the InfoMax  independent component analysis (ICA) algorithm, was used to identify and remove cardiac, ocular and muscular artefacts \cite{Mehrkanoon2016252, Mehrkanoon2014692, Mehrkanoon2014338, Mehrkanoon2014BC}. The first five minutes of artefact-free EEG data were discarded to avoid any transient (or non resting-state) brain activity and a four-minute epoch was selected from the remaining data for subsequent analysis.
\subsection*{Construction of the volume conduction and leadfield model of the infant head}
The volume conduction model refers to the EEG forward problem that estimates the electric potentials on the scalp for a known configuration of the electric dipole sources, provided that the physical properties of the head tissues such as conductivities are known \cite{Mehrkanoon2014BC}. 
Scalp EEG electrodes were first co-registered with the averaged normalised T$_2$-weighted MR image using a 9-parameter transformation of the EGI Hydrocell GSN 130 Geodesic Sensor Nets such that the electrodes were placed according to the 10-20 system \cite{Mehrkanoon2014BC}. Given the tissue priors of scalp, skull, cerebrospinal fluid (CSF), grey matter (GM) and white matter (WM) of the infant brain anatomical template \cite{ShiFeng2010}, the Matlab toolbox FieldTrip (http://www.fieldtriptoolbox.org/) was used to construct the realistic volume conduction model of the infant head by numerically solving the forward model \cite{DANG2011359}, which resulted in a leadfield matrix. The leadfield matrix is a mathematical operator that linearly transforms electrical currents generated at $N$ dipole sources in the brain into scalp potentials observed in EEG signals: $\mathbf{Y}(t)=\sum_{i=1}^3\mathbf{H}_i\mathbf{S}_i(t)+\mathbf{e}(t), \ i=x, y, z$, where $\mathbf{Y}(t)\in\mathbb{R}^{n_e\times \mathbf{T}}$ denotes the EEG signals (i.e., $n_e$ sensors/electrodes and $\mathbf{T}$ data points), $\mathbf{H}_{i=x}\in\mathbb{R}^{n_e\times N}$ denotes the leadfield matrix in the $x$ orientation, and hence $\mathbf{H}_y$ and $\mathbf{H}_z$ are the leadfield matrices in the $y$ and $z$ orientations respectively, $\mathbf{S}_i(t)\in\mathbb{R}^{N\times \mathbf{T}}$ is the dipole source, $\mathbf{e}(t)\in\mathbb{R}^{n_e\times \mathbf{T}}$ is the measurement noise which is assumed to have normal distribution with zero-mean and diagonal covariance matrix $\mathbf{\Sigma}_{\mathbf{ee}}$, i.e. $\mathbf{e}\sim\mathcal{N}(0,\mathbf{\Sigma}_{\mathbf{ee}})$. The conductivity values of the infant's head-component tissues were set to 0.43 $S/m$ (scalp), 0.2 $S/m$ (skull), 1.79 $S/m$ (CSF) and 0.3 $S/m$ (brain) \cite{Roche-Labarbe2008}. The locations of the dipole sources were then constrained by the infant GM volume with respect to predefined standard AAL neonatal brain template with 90 cortical and subcortical ROIs in the GM volume \cite{ShiFeng2010}. This process resulted in $N=$4,882 dipole locations in the 90 ROIs, each with three orientations in the $x$, $y$, and $z$ axis$-$ each dipole source represents 5mm$^3$ in volume. Thus, we used the leadfield array $\mathbf{H}\in\mathbb{R}^{(62\times 4,882)\times3}$ for electric source imaging in each of the 100 infants in subsequent analyses.
\subsection*{Electric source imaging of neonatal EEG}
We used linearly constrained minimum variance (LCMV) beamforming algorithm to estimate the magnitude of the dipole sources at the $i$th orientation \cite{BDVanVeen1997}:\\ $\hat{\mathbf{S}}_i(t)=(\mathbf{H}_i^T\mathbf{C}^{-1}\mathbf{H}_i)^{-1}\mathbf{H}_i^{T}\mathbf{C}^{-1}\mathbf{Y}(t)$, where $\mathbf{C}=\mathbb{E}[\mathbf{YY}^T]$ denotes the covariance matrix of sensor-space averaged-reference EEG signals, and $\mathbb{E}$ is the mathematical expectation. Note that the LCMV beamformer is robust to partial correlation between sources where the interdependencies of periodic neuronal population activity are preserved and that phase-synchronisation of interacting non-linear sources is not perturbed by the beamformer analysis \cite{Hipp2012}. From 4,882 seed voxels, we selected the seed voxels that are in the Euclidian center of each ROI in the left and right hemispheres, resulted in 90 seed-voxel signals, each with the three $x$, $y$, and $z$ components. Because the norm of each seed voxel time-series can lead to the occurrence of frequency doubling in the subsequent time-frequency analysis, we applied principal component analysis (PCA) to the covariance matrix of wide-band seed voxel time-series to determine a dominant orientation of the Euclidean centroid seed-voxel time-series \cite{Hipp2012, Mehrkanoon2014BC}. This process resulted in 90 time-series representing the neuronal population dynamics of each of the 90 ROIs.
\subsection*{Derivation of functional brain networks}
To quantify coherent spontaneous activity of spatially distributed neuronal populations in each neonatal brain, functional connectivity was measured by the imaginary part of the complex-valued time-frequency coherence (i.e., non zero-lag synchronisation) between all pairs of 90 ROI signals \eqref{eq:TFcoherence}. Taking the imaginary part of the coherence minimises the occurrence of spurious correlation among the nearby seed voxels due to the volume conduction property \cite{Nolte20042292, MehrkanoonEURASIP2013, Mehrkanoon2014338}. For $N$=90 ROIs, the maximum number of edges is given by $M=\frac{N\times(N-1)}{2}=4,005$ at each time-point and frequency for the $k$th neonatal brain, yielding functional connectivity array $\gamma_k(t,f)\in\mathbb{R}^{\mathbf{T}\times \mathbf{F}\times M}$:
\begin{equation}
\label{eq:S1}
\begin{split}
\gamma_k(t,f)=
\begin{bmatrix}
\Gamma_{(1,2)}(t,f) & \Gamma_{(1,3)}(t,f) & \Gamma_{(1,4)}(t,f) & \cdots & \Gamma_{(1,90)}(t,f) \\
\dagger & \Gamma_{(2,3)}(t,f) & \Gamma_{(2,4)}(t,f) & \cdots & \Gamma_{(2,90)}(t,f)\\
\dagger & \dagger & \Gamma_{(3,4)}(t,f) & \cdots & \Gamma_{(3,90)}(t,f)\\
\vdots & \vdots & \ddots & \vdots & \vdots \\
\dagger & \dagger & \cdots & \dagger & \Gamma_{(89,90)}(t,f)
\end{bmatrix},\\
k&=1,\cdots,100 \ , \\
t&=1,\cdots,\mathbf{T}, \\
f&=1,\cdots,\mathbf{F} \ ,
\end{split}
\end{equation}
where $\Gamma_{i,j}(t,f)$ is the imaginary part of the complex-valued time-frequency coherence between the ROI signals $x_i(t)$ and $x_j(t)$ computed by
\begin{equation}
\begin{split}
\label{eq:TFcoherence}
\Gamma_{ij}(t,f)&=\Im\Big(\frac{\mathbf{S}\left\{P_{ij}(t,f)\right\}}{\sqrt{\mathbf{S}\left\{P_{ii}(t,f)\right\}\mathbf{S}\left\{P_{jj}(t,f)\right\}}}\Big) \ i,j=1,\cdots,90, i\neq j,\\
P_{ij}(t,f)&=X_i(t,f)\tilde{X}_j(t,f), \ P_{ii}(t,f)= \left|X_i(t,f)\right|^2\\
X_i(t,f)&=x_i(t)\ast\psi(t,f),\\
&=x_i(t)\ast(a^2\pi)^{-1/4}\text{exp}(-\frac{t^2}{2a^2}-\mathbf{i}2\pi ft),
\end{split}
\end{equation}
where $\Im(.)$ denotes the imaginary operator, $P_{ij}(t,f)$ and $P_{ii}(t,f)$ are respectively the power and cross spectral densities, $\tilde{X}(t,f)$ is the complex-conjugate of $X(t,f)$, the asterisk $\ast$ is the convolution operator, $\psi(t,f)$ is the complex-valued Morlet wavelet, $a$ is the wavelet scale, $\mathbf{i}=\sqrt{-1}$, and $f$ is the wavelet center frequency \cite{MehrkanoonEURASIP2013}. The scale and frequency are proportionally related by $a=\frac{6}{2\pi f}$ \cite{MehrkanoonEURASIP2013}, and we chose a broad frequency spectrum ranging from 1 to 30 Hz by the increment of 0.5. The smoothing operator $\mathbf{S}\{.\}$ used in this study is defined by the Gaussian kernel $K(t,f)=\text{exp}\Big(-\Big(\frac{t^2}{2\sigma_t^2}+\frac{f^2}{2\sigma_f^2}\Big)\Big)$, where $\sigma_t=0.66 \  s$ and $\sigma_f=1.32$ Hz denote the time and frequency spreads of the kernel. The smoothing process was implemented by convolving the kernel $K(t,f)$ with the cross- and power-spectral densities to improve the reliability of the coherency estimate \cite{MehrkanoonEURASIP2013}. The symbol $\dagger$ denotes array entries that, due to the undirected nature of \eqref{eq:S1}, are symmetric across the main diagonal and these are not considered further. Since \eqref{eq:S1} is a lossless decomposition of the ROI signals, the elements in the array $\gamma_k(t,f)$ will inevitably be mutually correlated and contain considerable redundancies. That is, the spatiotemporal patterns and spectral contents of the brain FC structure are inter-related, generating folded spatiotemporal and spectral dynamics \cite{Mehrkanoon2014338}. To unfold the complex dynamics of the brain FC structure in each neonatal brain, multilinear PCA was used throughout the eigen-decomposition of the covariance matrix of the time-frequency array $\gamma_k(t,f)$. Since $\gamma_k(t,f)$ has three factors of time ($t$), frequency ($f$), and network edges ($M$=4,005), we reshaped $\gamma_k(t,f)$ into the 2-D array $\Theta_k(tf,l)\in\mathbb{R}^{[\mathbf{T}\times \mathbf{F}]\times M}$ as [time-frequency]-by-edges. The subject-level resting-state networks were then obtained from the eigen-decomposition of the covariance matrix of $\Theta_k(tf,l)$ as follows \cite{Mehrkanoon2014338}
\begin{equation}
\label{eq:mPCA}
\begin{split}
C_{\Theta\Theta_k}&=\mathbb{E}[\Theta_k^\text{T}\Theta_k]=V_kD_kV^\text{T}_k,\\
Y_k(m,tf)&=V^\text{T}_{{(1:m,:)}_k}\Theta^\text{T}_k \ \in\mathbb{R}^{m\times tf}, \ l=1,2,\cdots,M \ , \\
\end{split}
\end{equation}
where T denotes the matrix transposition, $k$ is the subject index, $V_k\in\mathbb{R}^{M\times M}$ is the orthogonal matrix whose eigenvector columns (or modes) correspond to the edges of the RSNs for the $k$th neonatal brain, $D_k$ is the diagonal matrix representing the eigenvalues of $C_{\Theta\Theta_k}$ such that $\text{diag}(D_k)=\{d_{1_k}>d_{2_k}>\cdots>d_{M_k}\}$ shows the variance of each evident RS mode. That is, each $M\times 1$ eigenvector contains the contributions of every ROI-pair combinations (or the edges of the network) to that specific eigenmode and hence reflects the spatial topology of each network. Notably, $M=4,005$, and hence each $M\times 1$ eigenvector can be easily reshaped into a 90$\times$90 FC matrix. The PCA was implemented by an efficient eigenvalue decomposition algorithm using Matlab function $\mathtt{eigs.m}$, and resulted in $m=5$ (out of 4,005) modes as being sufficient to span the phase space of resting-state dynamics for the subject-level network derivation \eqref{eq:mPCA}. The time-frequency principal components (PCs) associated with these five eigenmodes were then obtained from the  projection of the $k$th original functional-connectivity array $\Theta_k(tf,l)$ through five eigenmodes \eqref{eq:mPCA}, resulted in $Y_k(m,tf)$. The time-frequency projection array $Y_k(m,tf)$ contain the dynamics of RSNs in the $k$th infant brain. We focus on identifying between-infants (i.e. group-level) RSNs reliability and consistency. Therefore, the five eigenmodes obtained from each infant were spatially concatenated in the matrix $\mathbf{V}(n,l)=\Big[V_{(:,1:5)_1},\cdots,V_{(:,1:5)_{100}}\Big]^{\text{T}} \in\mathbb{R}^{100m\times M}, n=1,2,\cdots,100m, \ l=1,2,\cdots,4005$. A second PCA was computed to extract $p=5$ group-level PCA modes from the matrix $\mathbf{V}(n,l)$. The spatial independent component analysis (ICA)-SOBI (Second Order Blind Identification) of the five group-level PCA modes was computed to derive the independent RSNs that are commonly present in 100 infants \cite{Belouchrani554307, Mehrkanoon2014338, HusterReneJ20159}:
\begin{equation}
\label{eq:ICA}
\begin{split}
C_{\mathbf{VV}}&=\mathbb{E}[\mathbf{V}\mathbf{V}^{\text{T}}]=\mathbf{U}\mathbf{D}\mathbf{U}^\text{T},\\
\mathbf{Q}&=\mathbf{U}^{\text{T}}_{(1:p,:)}\mathbf{V}\in\mathbb{R}^{p\times M},\\
\mathbf{S_\text{ICA}}&=\mathbf{W}\mathbf{Q}\in\mathbb{R}^{p\times M}, \\
\end{split}
\end{equation}
where truncated $\mathbf{U}\in\mathbb{R}^{100m\times p}$ denotes the weighting matrix whose first five eigenvector columns contain the weights of the contribution of each infant's brain-networks (or mode) to a common group-level RSN space. The truncated $\text{diag}(\mathbf{D})=\{\mathbf{d}_1>\cdots>\mathbf{d}_{p}\}$ is the diagonal matrix representing the explained-variance (or scaling factors) of each of the five group-level PCA modes across 100 infants, $\mathbf{Q}\in\mathbb{R}^{p\times M}$ is a group-level PCA that maps the eigenmodes of individual infant's cortical activity onto a common space and $\mathbf{S}_{\text{ICA}}\in\mathbb{R}^{p\times M}$ denotes the $p=5$ independent RSNs. The unmixing matrix $\mathbf{W}\in\mathbb{R}^{p\times p}$ linearly decomposes the orthogonal group-level PCA modes $\mathbf{Q}$ into five independent RSNs denoted by $ \mathbf{S}_{\text{ICA}}$. Note that each row of the matrix $\mathbf{S}_{\text{ICA}}$ contains all possible network edges ($M=4,005$), and hence $\mathbf{S}_{\text{ICA}}$ was reorganised as five functional connectome matrices each with the size of $90\times90$; $\mathbf{S}_{\text{ICA}}=[\mathbf{s}_1,\cdots,\mathbf{s}_5]$, where $\mathbf{s}\in\mathbb{R}^{90\times 90}$. The group-level weighting matrix $\mathbf{U}$ can be shown as vertically augmented individual's weighting matrices $[\mathbf{u}_1,\cdots,\mathbf{u}_{100}]\in\mathbb{R}^{100\times(m\times p)}$. Therefore, the time-frequency characteristics of each of the 5 independent RSNs were obtained from mapping the time-frequency projection (or principal components) of the same $k$th infant, i.e. $Y_k(m,tf)$, onto group-level ICA space as follows:
\begin{equation}
\label{eq:tfICA}
\mathbf{Y}_{k}^{\text{ICA}}(p,tf)=\mathbf{W}\mathbf{u}^T_kY_k(m,tf) \ \in\mathbb{R}^{p\times tf},
\end{equation}
where $\mathbf{Y}_{k}^{\text{ICA}}(p,tf)$ denotes the time-frequency characteristics of the five independent RSNs in the $k$th infant in the group-level ICA space. For each infant, these five time-frequency spectra were reorganised as $\mathbf{y}_{i,k}(t,f), i=1,\cdots,p$, such that $\mathbf{y}_{i,k}(t,f)\in\mathbb{R}^{\mathbf{T}\times \mathbf{F}}$. We then used the time-frequency matrix $\mathbf{y}_{i,k}(t,f)$ for subsequent group-level time and frequency analysis of network dynamics across 100 infants.
\subsection*{Graph theoretical analysis}
To quantitatively evaluate the topological architecture of each of the five RSNs obtained from the group-level ICA, mathematical graph theory was used. First, each $90\times90$ RSN was thresholded with a connection density of 17\%; brain network analyses typically apply multiple thresholds centering around this value, which is typically employed in brain network research \cite{vandenHeuvel15775}. Two sets of graph connectivity measurements were then used to quantify the functional properties of these thresholded five RSNs (See below).
\subsubsection*{Primary connectivity metrics} A set of graph metrics computed for each of the five thresholded RSNs included \textit{node degree} ($K$) (the number of connections linked to each node), \textit{clustering coefficient} ($C$) (which quantifies the brain's ability in functional segregation or information processing; the number of triangular connections between the nearest node neighbours of a node divided by the maximal possible number of such connections \cite{WattsDuncan1998}, \textit{betweenness centrality} (a measure of a node hubness; a fraction of the shortest paths between all other node pairs in the network that pass through any particular node), \textit{global efficiency} ($\mathbf{E}_{\text{glob}}$) (a measure of the brain's ability in functional integration or binding the information processed by the brain regions \cite{WattsDuncan1998}, \textit{path length} ($L$) (the minimal number of edges that must be traversed to form a direct connection between the two nodes of interest in the network \cite{WattsDuncan1998}, which is related to a measure of the brain functional integration by averaging the shortest path lengths between all pairs of nodes in the network), \textit{assortativity coefficient} (a measure of the tendency of high-degree nodes to connect with each other, which is a relative measure of the resilience (or elasticity) of the network against sudden attacks to the network components such as hub nodes and edges in the network \cite{Achard200663}).
\subsubsection*{Complex connectivity analysis}
\textit{Small-world} topology: We identified the small-worldness of each of the five RSNs by computing the ratio of the normalised clustering coefficient, $\gamma=C/C_\text{random}$, to the normalised path length, $\lambda=L/L_{\text{random}}$. $C_\text{random}$ and $L_\text{random}$ respectively denote the average clustering-coefficient and average path-length of the surrogate networks constructed by randomising all connections of the original RSN under analysis (10,000 realisations) while preserving the dimension and degree distribution of the original network. In a small-world network, we expect the ratio $\lambda=L/L_{\text{random}}$ tends to 1, and the ratio $\gamma=C/C_\text{random}$ to be greater than 1, and hence the small-world index defined as $\sigma=\gamma/\lambda$ is greater than 1 \cite{WattsDuncan1998} $-$ indicative of a dense local clustering (or cliquishness) between neighbouring nodes yet a short path length between any distant pair of nodes due to the existence of relatively few long-range connections.\\
\textit{Rich-club organisation}: The centrepiece of the graph theoretical analysis of the functional connectomes in this paper is the exploration of the \textit{rich-clubs} of functional hub nodes in the infants' RSNs by computing the \textit{rich-club coefficient} $\phi(K)$ across a range of node degrees $K$. Rich-club organisation implies that the hub nodes of a network are more strongly connected with each other than expected by their high degree $K$ alone \cite{Colizza2006110, Schroeter20155459}. For each neonatal RSNs, a weighted rich-club coefficient $\phi^w(K)$, at each degree level of $K$, was computed as the ratio of the sum of the edge weights that connect a group of nodes (the club) of degree $\geqslant K$ to the sum of the strongest edge weights that such nodes could share with other nodes in the whole network \cite{vandenHeuvel15775}:
\begin{equation}
\label{eq:RC}
\phi^W(K)=\frac{W_{>K}}{\sum_{l=1}^{E_{>K}}W_l^{\text{ranked}}}, l=1,2,\cdots, E\ ,
\end{equation}
where $W_{>K}$ denotes the sum of the weights of the number of edges between the members of the club with a degree of $\geqslant K$ given by the strongest $E_{>K}$ edges of the network, the $E$ is the total number of edges, and $W_l^{\text{ranked}}\geqslant W_{l+1}^{\text{ranked}}\geqslant\cdots\geqslant W_{E}^{\text{ranked}}$ is a series representing the ranked weights on all connections of the network. A network will exhibit a rich-club topological organisation if, for a range of degree $K$, the normalised weighted rich-club coefficient, $\phi_\text{norm}^W(K)=\frac{\phi^W(K)}{\Phi^{W}_{\text{null}}(K)}$, is $>1$ \cite{vandenHeuvel15775, Schroeter20155459}. The $\Phi^{W}_{\text{null}}(K)$ is the rich-club effect assessed on the appropriate surrogate (null) model constructed by randomising the network edges (10,000 realisations, see above). For each RSN, 10,000 randomised networks were first constructed. The weighted rich-club coefficient $\phi^{W}_{\text{null}}(K)$ was then computed for each level of degree $K$, for each randomised network. The average rich-club coefficient over the 10,000 randomised networks was then used as $\Phi^{W}_{\text{null}}(K)$ to compute the normalised weighted rich-club coefficient $\phi_\text{norm}^W(K)$. 
\subsubsection*{Statistical analysis of rich-club organisation} Permutation testing was used to assess statistical significance of the rich-club organisation of each neonatal RSN. The null distribution of rich-club coefficients $\phi^{W}_{\text{null}}(K)$ were first obtained from the 10,000 random networks (see above). For each degree $K$ that $\phi_\text{norm}^W(K)>1$, we tested whether the rich-club coefficient $\phi_\text{norm}^W(K)$ is significantly greater than the 95\% of the null $\phi^{W}_{\text{null}}(K)$ \cite{vandenHeuvel15775}. The null distribution was rank ordered and the 95\% confidence interval determined non-parametrically from this rank order; ($p<0.05$, one-tailed test). The rich-club coefficients $\phi_\text{norm}^W(K)$ lies above this threshold were considered statistically robust rich-club regime.
\subsection*{Dynamics of RSNs in infants}
The exploration of the spectral fingerprints and temporal dynamics of neonatal RSNs was performed by analysing the time-frequency spectra (i.e. $\mathbf{y}_{i,k}(t,f), i=1,\cdots,5$) of each of the five RSNs for the $k$th infant. The spectral fingerprints of a network reflect the frequency range at which the network of interest oscillates \cite{Mehrkanoon2014338}. Temporal dynamics of each network quantifies the variability of the network of interest over time \cite{Mehrkanoon2014338, Vidaurre12827}. 
\subsubsection*{Spectral fingerprints and temporal dynamics of RSNs}
The spectral property and the temporal fluctuations of the networks were independently investigated by analysing the marginal densities of the group-level time-frequency spectra $\mathbf{y}_{i,k}(t,f)$ \cite{Mehrkanoon2014338}. The frequency spectrum of each RSN is given by $y_{i,k}(f)=\frac{1}{\mathbf{T}}\sum_{t=0}^{\mathbf{T}}\mathbf{y}_{i,k}(t,f)$, representing the \textit{fast time-scale} fluctuations of each RSN. The spectral fingerprints of each of the five RSNs were computed by the grand average and standard deviation of $y_{i,k}(f)$ across 100 infants. The temporal dynamics of RSNs within each infant were given by the instantaneous-mean of $\mathbf{y}_{i,k}(t,f)$ as $x_{i,k}(t)=\frac{1}{\mathbf{F}}\sum_{f=1}^{\mathbf{F}}\mathbf{y}_{i,k}(t,f)$. The group-level temporal dynamics of each RSN energy underpinning each RSN were computed by the grand-average and standard-deviation of the instantaneous-mean dynamics, $x_{i,k}(t)$, of each RSN across 100 infants. The power spectral density of the temporal dynamics ($x_{k,m}(t)$) of each RSN was computed to infer slow modulations of fast oscillatory activity of the network, which is an estimation of the envelope of the fast oscillatory activity of RS cortical network \cite{Mehrkanoon2014338}.
\subsubsection*{Long-range dependencies of RSN dynamics}The behaviour of a complex dynamical system over time is ``sensitive dependence on an initial condition'' of the system. That is, the past behaviour of the system influences its future behaviour, indicating a long-range dependence property of the process that governs the system dynamics, which is related to fractality \cite{Kobayashi1999159}. The Hurst exponent is a measure of the extent of long-range dependence of such process that is reflected by time-series obtained from the system of interest. Here, for each infant, the behaviours of RS cortical networks are reflected by their affiliated temporal dynamics ($x_{i,k}(t)$, see above). Therefore, we further assessed the temporal dynamics of the networks by computing the Hurst exponent of each network's temporal fluctuations as $H_{i,k}=\frac{\text{log}(R_{i,k}/S_{i,k})}{\text{log}(T)}$, where $R_{i,k}/S_{i,k}$ denotes the so-called range statistics: $R_{i,k}$ is the difference between the maximum and minimum deviation from the mean, and $S_{i,k}$ is the standard deviation over total number of $T$ samples for the $i$th network in the $k$th infant, i.e., $0<H_{i,k}<0.5$ indicates short-range dependence, $H_{i,k}=0.5$ indicates a random-walk process that is uncorrelated, and $0.5<H_{i,k}<1$ indicates the existence of long-range dependencies or persistence \cite{Kobayashi1999159}. The Hurst exponent of the temporal dynamics $x_{i,k}(t)$ of each network expressed is given by
\begin{equation}
\label{eq:HurstExpo}
\frac{R_{i,k}}{S_{i,k}}=\frac{\underset{1\leq J\leq \mathbf{T}}{\text{max}}\vartheta_{i,k}-\underset{1\leq J\leq\mathbf{T}}{\text{min}}\vartheta_{i,k}}
{\sqrt{\frac{1}{\mathbf{T}}\sum_{t=1}^{\mathbf{T}}(x_{i,k}(t)-\mu_{x_{i,k}})^2}} \ , i=1,\cdots,5; k=1,\cdots,100 \ ,  
\end{equation}
where $\vartheta_{i,k}=\sum_{t_0=1}^{J}x_{i,k}(t_0)-\mu_{x_{i,k}}$, $\mu_{x_{i,k}}=\frac{1}{\mathbf{T}}\sum_{t=1}^{\mathbf{T}}x_{i,k}(t)$, $\vartheta_{i,k}$ denotes the deviation of $i$th network's temporal dynamics from its mean value $\mu_{x_{i,k}}$, and $J=\mathbf{T}/4$ is a non-overlapping sample size.
\subsubsection*{Temporal organisation of interacting RSNs}
Because these small number of RSNs capture the dynamics of the neonatal brain states in a low-dimensional orthogonal space, we sought to investigate \textit{between-networks temporal interactions}. Such interactions or transitions between the brain states that are captured by a set of networks are referred to as \textit{meta-state} dynamics \cite{Vidaurre12827}. To identify the pattern of interactions between the five RSNs, we assess the cross-correlation between the temporal dynamics of the networks $\bar{x}_{i,k}(t)$. The cross-correlation between the temporal dynamics of two networks of $m$ and $m'$ in the $k$th infant is given as a function of a time-lag $\tau$ by
\begin{equation}
\begin{split}
\label{eq:xcorr}
r_{(m,m')_k}(\tau)&=\mathbb{E}\{\bar{x}_m(t)\bar{x}_{m'}(t+\tau)\}, \ m,m'=1,\cdots,5, \ m\neq m',\\ \tau&=-\mathbf{T},\cdots,\mathbf{T}.
\end{split}
\end{equation}
To extract the component of the cross-correlation functions that is not invariant to time reversal, we determine the asymmetric components of $r_{(m,m')_k}(\tau)$ as follows
\begin{equation}
\label{eq:asymmetric}
\tilde{r}_{(m,m')_k}(\tau)=\frac{r_{(m,m')_k}(\tau)-r^{\text{T}}_{(m,m')_k}(\tau)}{2}.
\end{equation}
The slope signs of the asymmetric component near the zero time-lag, $\tilde{r}_{(m,m')_k}(0)$, were used to represent the relative leader (positive sign) and follower (negative sign) positions of the networks $m$ and $m'$. Examining all pairs of networks in each infant allows a temporal rank order (sequence) to be obtained. Confidence intervals for the asymmetric component of the cross-correlation were estimated using a non-parametric permutation approach. To this end, 1,000 surrogate time-series were constructed by random permutation of the network times-series. The asymmetric component of the cross-correlation functions \eqref{eq:xcorr} derived from these surrogate time-series represent trivial non-zero fluctuations of the magnitude expected from data of that string length and amplitude distribution but with no further structure of interest. Two-sided 95\% confidence intervals were estimated by rank ordering this null distribution and choosing the 25$^{th}$ and 975$^{th}$ values. This process was repeated independently for each of the five networks of all infants, and the signs of the grand average asymmetric component were used to determine the temporal sequence (or cycle) between the five robust RSNs. The time-lags that the networks need to take their leading/following positions during network interactions were determined by the first time point at within the network crosses $t=0$ s. These time-lags were then used to distinguish between the lagged-expression of the networks, yielding the patterns of cyclic temporal interactions of the networks.

\section*{Results}
\subsection*{Anatomically constrained brain networks inferred from EEG source estimates in infants}
We recorded 64-channels EEG from 100 infants in eyes-closed active-sleep condition (average duration of $\sim$ 240 s), followed by artefact rejection and EEG source reconstruction on a realistic volume conduction model of the infant head (Fig.1\textbf{\textbf{A-B}}) in order to estimate the magnitudes of spatially distributed neuronal populations activity in the pre-defined Automated Anatomical Labelling (AAL) neonate template into 90 cortical and sub-cortical grey matter (GM) regions of interest (ROIs) \cite{ShiFeng2010} (Fig.1\textbf{\textbf{A-B}}). The group-level independent component analysis (gICA) of the subject-level multilinear principal component analysis (mPCA) of non-zero time-lagged time-frequency coherence between the neuronal populations signals in these 90 ROIs revealed spatially, spectrally and temporally resolved distinct modes of functional brain networks that are consistently expressed across 100 infants (Jackknifing test, one-sided $P<$0.01) (Fig.1\textbf{\textbf{B}} right panel, \textbf{\textbf{C-D}}). The power spectral density of the neuronal populations signals in these 90 anatomical ROIs shows statistically significant oscillations from the delta to low alpha frequency ranges (1-10 Hz) (one-sided $P<0.01$, z-score=2.5) dominated by low-frequency ranges (1.5-5 Hz) in both hemispheres (Fig.1\textbf{\textbf{E}}).\\
\begin{figure*}[htbp]
{\includegraphics[height=18cm,width=14cm]{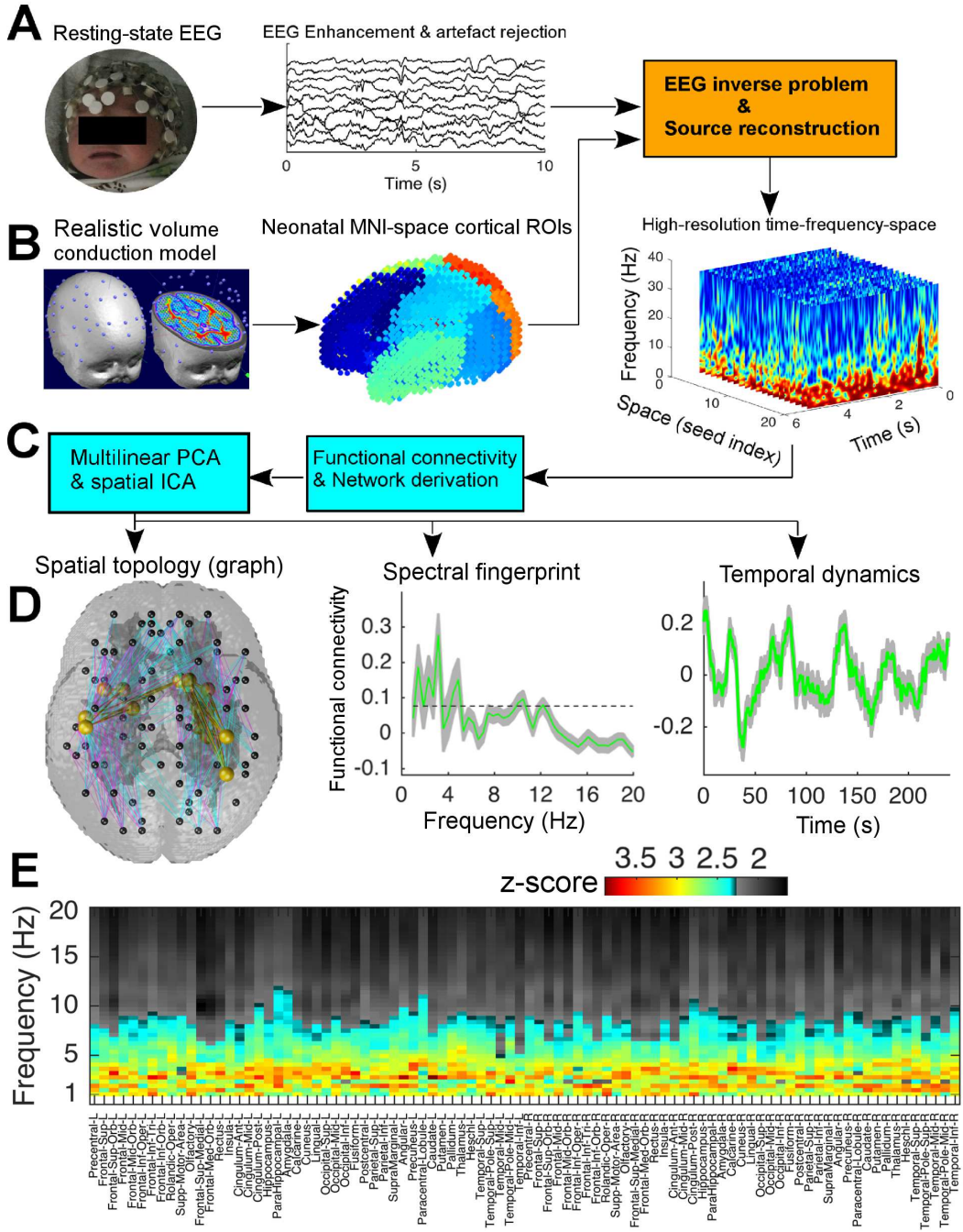}}
\caption{
\textbf{Construction of large-scale brain network dynamics in infants at term age.} \textbf{A,} EEG acquisition, artefact correction and beamforming-based source estimation of neuronal populations activity using, \textbf{B}, construction of MNI-space volume conduction model of the head including 90 ROIs in the grey-matter volume of the infant brain. \textbf{C,} network derivation using group-level ICA of subject-level PCA-driven time-frequency functional networks, yielding \textbf{D}, spatial architectures, spectral fingerprints and temporal dynamics of the networks. \textbf{E}, z-score statistics of the group-level power spectra of neuronal populations activity in the 90 ROIs.
}
\label{F1}
\end{figure*}
\subsection*{Rich-club organisations reflect cortical network states in infants}
Our network derivation approach revealed five spatially independent patterns of RSNs, each with distinct rich-club curves associated with  specific rich-club topological organisation (Fig.2). The rich-club of interconnected cortical hubs in these five RSNs respectively show the existence of the default mode network (DMN), primary-sensory limb system network, thalamo-frontal network, thalamo-sensorimotor and visual-limbic system networks (Fig.2). The normalised weighted rich-club curve (red), as the ratio of the weighted rich-club curve ($\phi^W(K)$, blue) to the rich-club of random network ($\phi_\text{rand}^W(K)$, grey), is significantly greater than 1 (10,000 permutation tests, one-sided $P<0.01$) for the five RSNs in a set of node-degree ranges: $27\leq K\leq35$ (DMN), $22\leq K\leq31$ (primary-sensory limb system network), $24\leq K\leq34$ (thalamo-frontal network), $23\leq K\leq$33 (thalamo-sensorimotor network), $22\leq K\leq$31 (visual-limbic system network) (Fig.2, left column).
\vspace{2mm}\\
\textbf{The default mode network.}
The rich-club organisation of the DMN at the node-degree $K=31$ spans the left and right cortical hubs (gold) that are connected with each other by the rich-edges (gold): Right superior frontal lobe, homologous posterior cingulate cortices, left cuneus and right precuneus in the basic visual processing system (Fig.2\textbf{\textbf{A}} (glass-brain panels), Fig.3\textbf{\textbf{A}}). The functional connections between the basal ganglia (pallidum and caudate $-$ the limbic system), right superior frontal lobe and posterior cingulate cortex are richly structured in the DMN throughout the homologous thalamus regions (Fig.3\textbf{\textbf{A}}). The non-hub cortical regions of homologous occipital lobes, frontal mid and superior gyri, left supplementary motor area and somatosensory cortex, have a broad range of degrees $1\leq K\leq30$, where $K=1$ (dark purple) and $K=30$ (bright purple), which are densely connected to cortical hubs throughout the feeders (cyan) (Fig.2\textbf{\textbf{A}}, glass-brain panels, Fig.3\textbf{\textbf{A}}).
\vspace{1mm}\\
\textbf{Primary sensory$-$limbic system network.}
The rich-club architecture of the second network at the degree $K=27$ reveals the primary functional hubs, mainly inter-hemisphericly and bilaterally, including the primary auditory cortex (Heschl), primary somatosensory cortex (left postcentral), left insula (involved in the motor control and cognitive functioning), inferior parietal lobe, right olfactory, Rolandic operculum, right hippocampus, amygdala and basal ganglia (putamen and caudate) (Fig.2\textbf{\textbf{B}} (glass-brain panels), Fig.3\textbf{\textbf{B}}). The topological organisation of the feeders densely links the non-hub cortical regions of superior, middle and inferior temporal, frontal and occipital lobes ($1\leq K\leq25$) with the rich-club network within and between the hemispheres throughout the limbic system, somatosensory and the primary auditory cortex (Fig.3\textbf{\textbf{B}}).
\vspace{2mm}\\
\textbf{Thalamo-frontal network.}
The rich-club of hub cliques in the third network spans the limbic system and frontal gyri including the homologous thalamus regions, posterior and anterior cingulate cortices, right hippocampus, superior, medial and middle frontal orbits and rectus at the degree $K=28$ (Fig.2\textbf{\textbf{C}} glass-brain panels, Fig.3\textbf{\textbf{C}}). The feeders link non-hub cortical regions of the right supplementary-motor area, left hippocampus, parahippocampal, amygdala, homologous olfactory systems, basal ganglia (putamen, pallidum and caudate) and occipital lobe (in a range of degrees $1\leq K\leq26$), with the cortical hub regions of homologous anterior cingulate cortices and orbitofrontal gyri in the network (Fig.3\textbf{\textbf{C}}).
\vspace{2mm}\\
\textbf{Thalamo-sensorimotor network.}
The spatial pattern of the rich-club of hubs in the fourth RSN reveals the sensorimotor-thalamic network spanning the thalamus, basal ganglia (putamen and caudate), supplementary motor area, primary motor cortex, somatosensory, superior/inferior frontal gyri at degree $K=27$ (Fig.3\textbf{\textbf{D}} (glass-brain panels), Fig.3\textbf{\textbf{D}}). The feeders densely link the non-cortical hubs of occipital lobes, superior, middle and inferior fronto-parietal lobes, and cingulate cortices, in a range of degrees $1\leq K\leq25$, with the rich-club of cortical hubs. The temporal lobe and primary auditory cortices have minimal functional connections to the rich-club of hubs compared to the other non-hub regions.
\vspace{2mm}\\
\textbf{Visual-limbic system network.}
The spatial pattern of the rich-club of hub cliques in the fifth RSN shows the visual-limbic system network spanning the superior/middle occipital gyri, cuneus, putamen, posterior cingulate cortex, parahippocampal gyrus, hippocampus, amygdala and olfactory at $K=27$ (Fig.2\textbf{\textbf{E}} glass-brain panels, Fig.4). The feeders, throughout the rich-club of occipital lobe and the limbic system, link the homologous frontal lobes and anterior/middle cingulate gyri and the right occipital lobe (bright purple nodes) with the rich-club brain hubs in a range of degrees $1\leq K\leq25$. Interestingly, the rich-club of hubs are more prominent in the left visual cortex, cuneus and putamen than the homologous regions in the left hemisphere.
\newpage
\begin{figure*}
\centering
{\includegraphics[height=16cm, width=14cm]{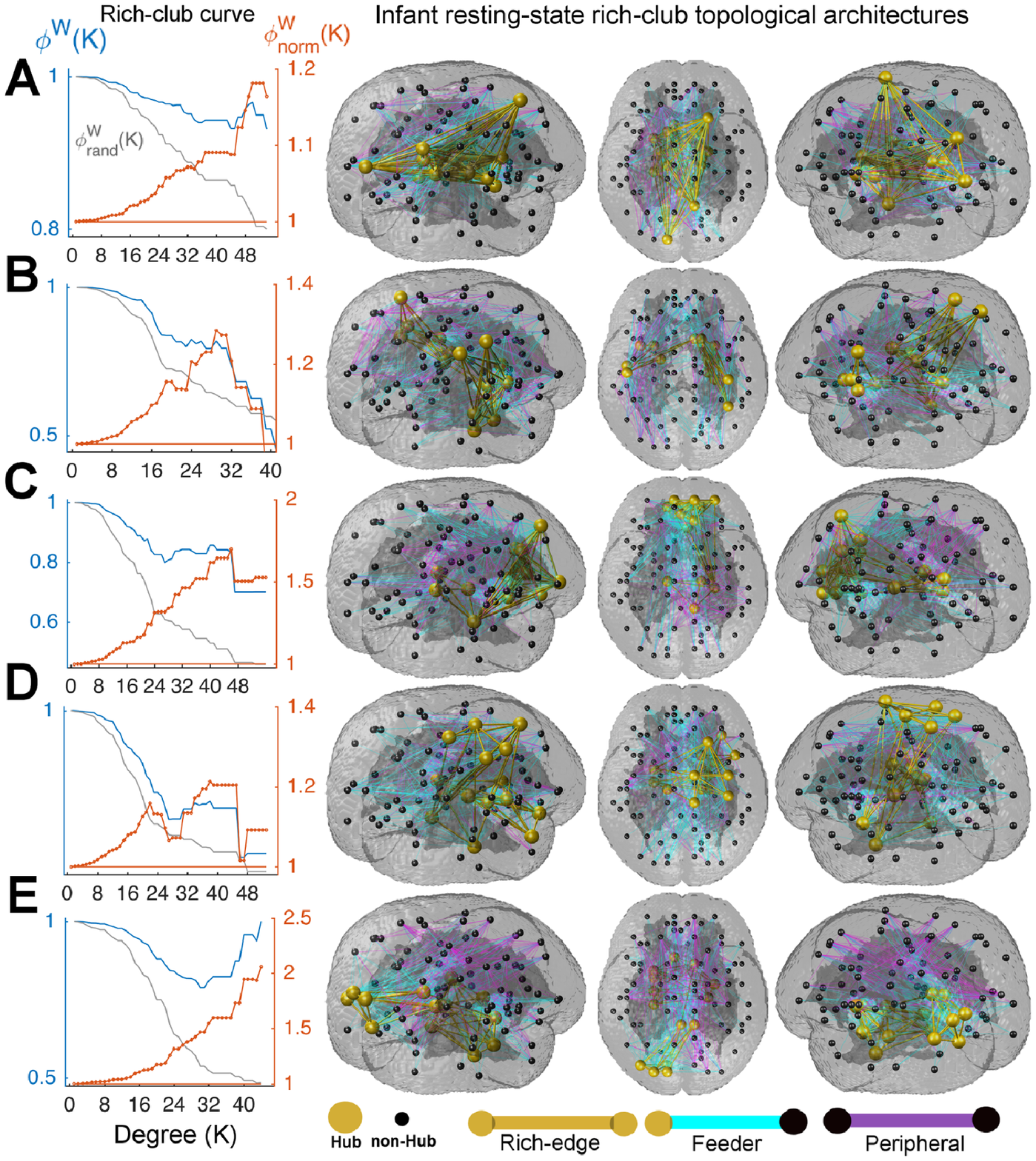}}
\caption{
\textbf{Rich-club coefficients (left column) and organisation of RSNs (glass-brain panels).} \textbf{(A-E,} left column) show the measures of weighted- ($\Phi^W(K)$, blue), null- ($\Phi_{\text{rand}}^W(K)$ , grey) and normalised ($\Phi_{\text{norm}}^W(K)$, red) rich-club curves for the DMN (that $\Phi_{\text{norm}}^W(K)$ significantly increases in a range of degrees $27\leq K\leq35$), primary-sensory limbic system network ($22\leq K\leq31$), thalamo-frontal ($24\leq K\leq34$), thalamo-sensorimotor ($23\leq K\leq33$), visual-limbic system ($22\leq K\leq31$). \textbf{A-E,} glass-brain panels): show rich-club of cortical hubs (gold) that are connected by the rich-edges (gold). The feeders (cyan) and peripheral edges (purple) respectively connect the non-hub regions to cortical hubs and other non-hubs.
}
\label{F2}
\end{figure*}
\begin{figure*}
\centering
{\includegraphics[height=16cm, width=14cm]{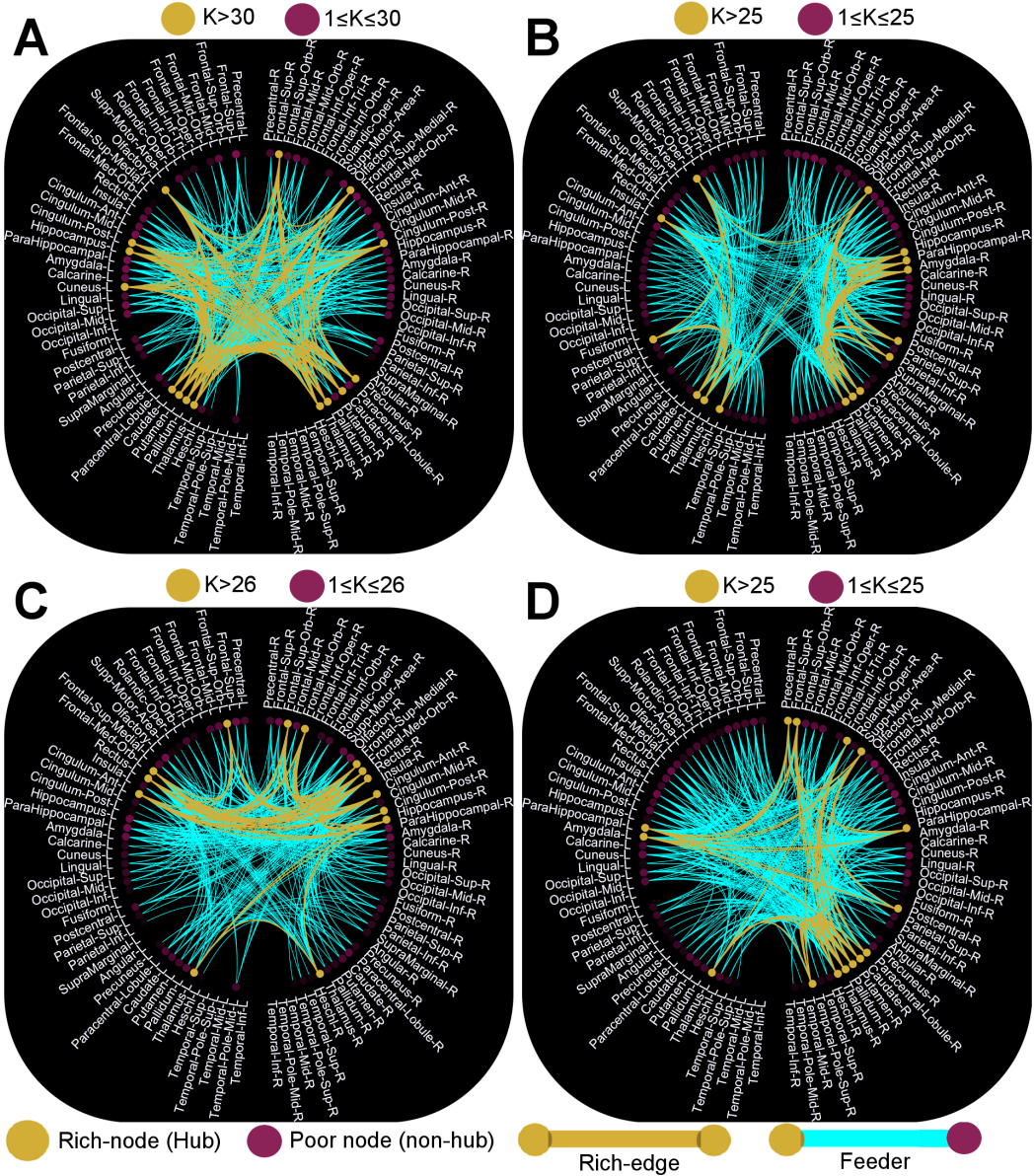}}
\caption{
\textbf{Functional connectome of the rich-club organisation of the first four RSNs.} \textbf{A}, the DMN with rich-club of 15 cortical hubs (gold) connected by the the rich edges (gold) at degree $K=31$. The feeders (cyan) connect cortical hubs to non-hub regions in a range of degrees $1\leq K\leq30$ (purple), where $K=1$ (dark purple) and $K=30$ (bright purple). \textbf{B,} the primary sensory-limbic system network: Inter-hemispherically dominant rich-club of 15 cortical hubs at $K=27$, where non-hub cortical regions with inter-hemispheric feeders have a range of degrees $1\leq K\leq25$. \textbf{C,} thalamo-frontal network: Bilaterally organised rich-club of 14 cortical hubs at $K=28$, where non-hub regions with intra-hemispheric feeders have a range of degrees $1\leq K\leq26$. \textbf{D,} thalamo-sensorimotor network: Right-hemispherically dominant rich-club of 15 cortical hubs at $K=27$, where the non-hub regions with bilateral feeders are confined in the degree intervals of $1\leq K\leq25$.
}
\label{F3}
\end{figure*}
\begin{figure*}
{\includegraphics[width=0.45\textwidth]{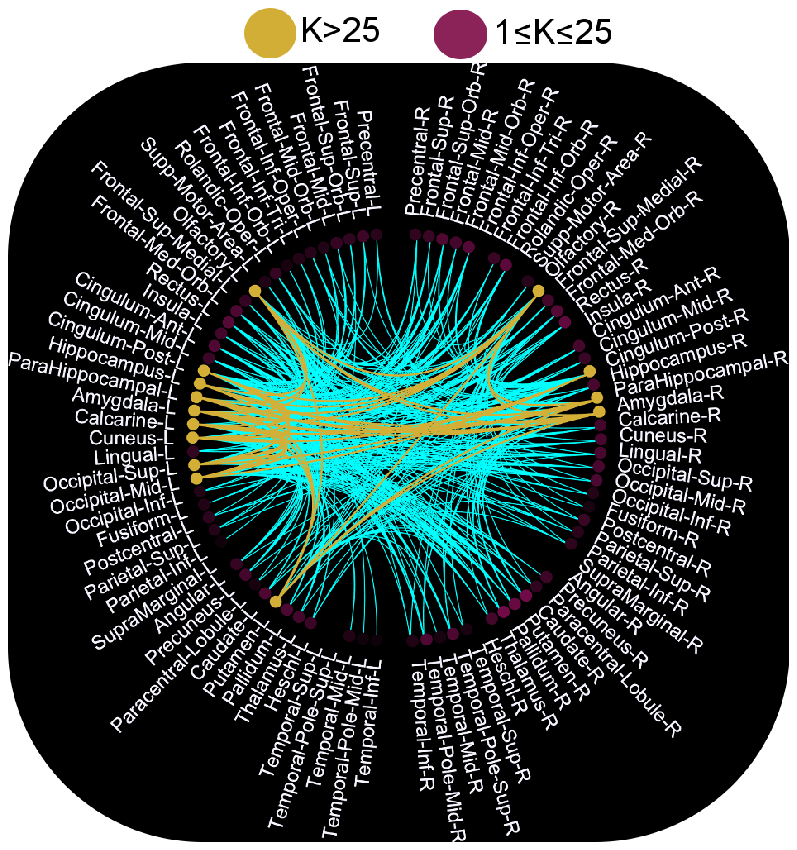}}
\caption{
\textbf{Functional connectome of the rich-club organisation of visual-limbic system network.} The left-hemispherically dominant rich-club of 14 cortical hubs at $K=27$, and the non-hub regions, mainly bilateral feeders, in a range of degrees $1\leq K\leq25$. The non-hubs of homologous frontal lobes and anterior/middle cingulate gyri and the right occipital lobe (bright purple nodes) are linked with the rich-club of cortical hubs of in a range of occipital lobe and the limbic system.
}
\label{F4}
\end{figure*}
\subsection*{Richly structured RSNs in infants are functionally vulnerable}
Graph theoretical analysis of the networks revealed the co-existence of functional integration and segregation capacity, small-world topology and functional vulnerability in each of the five spatially independent RSNs (Table 1). Although the global node-degree/strength of the DMN (31/25) is greater than the others confined in the intervals of [27/13, 29/16], the visual-limbic system network shows a greater mean betweenness-centrality (282.14) than the others nested in a range of [128, 278]. All networks show high mean clustering-coefficient and small short-path-length in the intervals of  [65, 78] and [0.6, 1.07] respectively. These five RSNs exhibit small-world topology with the values significantly greater than 1 (1,000 network randomisation tests, one-sided $P<0.01$), which are bounded in the intervals of [1.15, 1.35] (Table 1). Interestingly, although the normalised values of the global efficiency measures of the networks are greater than 1 (1,000 network randomisation tests) that indicates the existence of functional integration capacity in the infant brain at term age, the negative values of the assortativity coefficients of the networks (i.e., [-0.52, -0.35]) reveal the vulnerability (or weak resilience) of high-degree hubs in these five RSNs \cite{Maslov2002910}.
\begin{table}[htbp]
{
  \small
  {\bf Table 1} $|$ Graph theoretical metrics of resting-state networks in infants at term-age. 
}
\begin{center}
  \resizebox{.95\textwidth}{!}{%
  \begin{tabular}{@{} cccccc @{}}
    \toprule
    Graph metrics & Default mode & Sensory-limbic & Thalamo-frontal & Thalamo-senosrimotor &Visual-limbic \\ 
    \midrule
    Degree/strength$^a$ & 31/25 & 27.3/15.25 & 28.7/13.8 & 27.8/15.9 & 27/15.92 \\
    Clustering coefficient & 76.66 & 68 & 76.01 & 65.17 & 77.54 \\ 
    Betweenness & 128.9 & 277.9 & 201.6 & 245.35 & 282.14 \\ 
    Global efficiency$^b$ & 1.011 & 1.021 & 1.011 & 1.021 & 1.01 \\ 
    Assortativity & -0.47 & -0.52 & -0.35 & -0.44 & -0.45 \\ 
    Average path length & 1.07 & 0.68 & 0.72 & 0.86 & 0.83\\ 
    Small-world index & 1.25 & 1.28 & 1.34 & 1.15 & 1.26 \\
    \bottomrule
  $^a$\text{Measures represent mean + I SD}.\\
  $^b$\text{Normalised values.}\\
  \end{tabular}}
  \end{center}
  \label{tab:graph metrics}
\end{table}
\vspace{-15mm}\\
\subsection*{Spectral fingerprints of RSNs in infants reflect the already established cortical network-dependent carrier frequency in infants at term age}
The spectral fingerprints of the five RSNs reveal the carrier frequency (or fast timescale) at which the spatially distributed neuronal populations in the brain exhibit synchronous regime over time \cite{Mehrkanoon2014338, Hipp2012}. In the DMN, a continuous range of significant functional connectivities (0.28-0.57, one-sided $P<0.01$, z-score=2.45) occur in the delta and low theta bands (1-5.5 Hz) with small standard devision across infants (red, Fig.5\textbf{\textbf{A}} left panel), whereas the primary-sensory limbic system network exhibits significant functional connectivity (0.08-0.28) at multiple discrete frequencies with the peaks at 1.5 Hz, 3 Hz, 4.5 Hz and 11 Hz, which have a larger standard deviation across infants (green, Fig.5\textbf{\textbf{B}} left panel). The fast timescale of the thalamo-frontal network demonstrates significantly strong functional connectivity (0.15-0.4, one-sided $P<0.01$, z-score=2.5) with three discrete peaks at 2 Hz and 3.5 Hz (the delta band) and 6 Hz in the theta band (blue, Fig.5\textbf{\textbf{C}}  left panel). The spectral fingerprints of the sensorimotor-thalamic and visual-limbic system networks respectively show significantly strong functional connectivity values of 0.38 (cyan) and 0.32 (pink) (one-sided $P<0.01$, z-score=2.4) at 1.6 Hz (low delta band) followed by multiple peaks in the frequency intervals of [2.5-6] Hz (Fig.5\textbf{\textbf{D}}), [2-3] Hz and 10 Hz (Fig.5\textbf{\textbf{E}}). Interestingly, the fast timescales of the networks of primary-sensory limbic system (Fig.5\textbf{\textbf{B}}), sensorimotor-thalamic (Fig.5\textbf{\textbf{D}}) and the visual-limbic system (Fig.5\textbf{\textbf{E}}) show the existence of a significant 10 Hz component in their spectral fingerprints (one-sided $P<0.01$, z-score=2.4).
\begin{figure*}
\centering
{\includegraphics[width=0.7\textwidth]{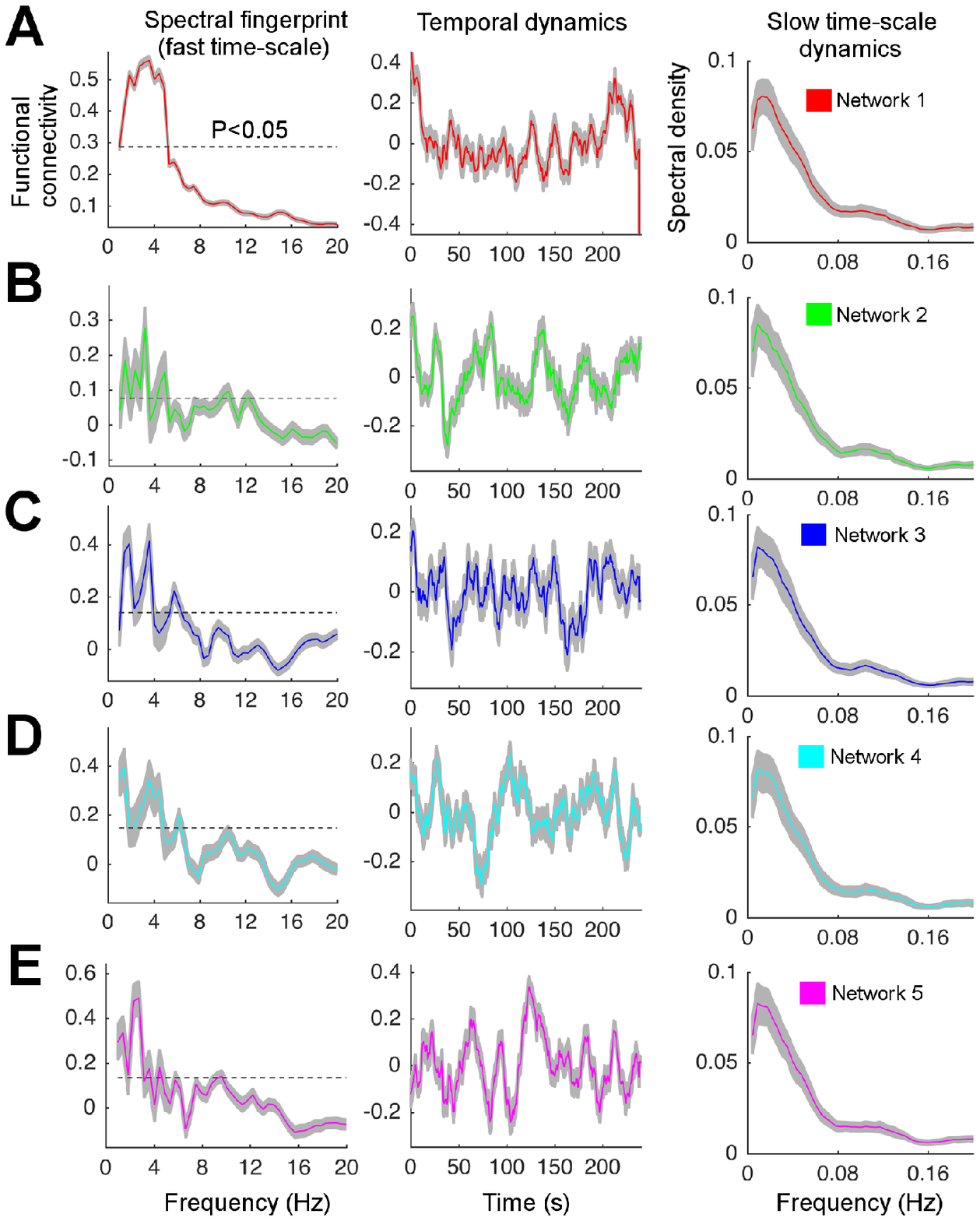}}
\caption{
\textbf{Spectral fingerprints, temporal dynamics and slow timescale oscillations of the five robust RSNs in 100 infants.} \textbf{A-E}, Left column: Neuronal populations synchronisation profile of RSNs captured by their frequency spectra over the time intervals of [0, 240s]. \textbf{A-E}, Middle column: Instantaneous fluctuation profile of RSNs inferred from their temporal dynamics over the time intervals of [0, 240s]. \textbf{A-E}, Right column: Nested-dynamics profile of RSNs obtained from their envelopes of the temporal dynamics.
}
\label{F5}
\end{figure*}
\subsection*{Infant brain networks exhibit richly organised temporal dynamics}
The instantaneous fluctuations of the fast timescale of the RSNs are reflected by the temporal dynamics at slow timescales, unfolding the brain network dynamics \cite{Mehrkanoon2014338}. The temporal dynamics of the five RSNs show slow timescale oscillatory patterns of the connectivity strengths confined in the intervals of [-0.3, 0.3] with multiple peaks between $20s\leq t\leq220s$ (Fig.5 middle column). The DMN largely fluctuates in the time intervals of $[100, 220]s$ (red, Fig.5\textbf{\textbf{A}} middle column), and the temporal dynamics of other four RSNs show multiple fluctuations across a broader time intervals of $[10, 240]s$ (green, blue, cyan and pink) respectively (Fig.5 \textbf{\textbf{B-E}} middle column). Together, all RSNs' temporal dynamics show small standard-deviation across infants ($<0.1$) (grey curves, middle column, Fig.5 \textbf{\textbf{B-E}} middle column). The power spectral density of the temporal dynamics of these RSNs show 1/$f$-like distribution that is quite stationary across the five networks, revealing the frequency of slow timescales of the dFC in infants (Fig.5\textbf{\textbf{A-E}} right column). Remarkably, temporal dynamics of all RSNs exhibit highest power at frequencies well below 0.16 Hz, which is the characteristic frequency of RS-fcMRI (Fig.5 right column). All RSNs show peaks at $f=0.022$ Hz and $f=0.12$ Hz respectively.
\newpage
\subsubsection*{Long-range correlations of temporal network dynamics reflect scale invariance structure of cortical activity in infancy.}
Because the human brain can be thought of as a complex dynamical system \cite{Kobayashi1999159, Mehrkanoon2014338}, temporal activity of such system is ``sensitive dependence on an initial condition'' of the system \cite{0951-7715-6-6-014}, indicating that the past temporal activity of the brain influences its future behaviour \cite{Kobayashi1999159}. To quantify the extend of a long-range dependence property of the infant brain network dynamics, the Hurst exponent analysis of the temporal dynamics of the five RSNs that exhibit apparent power-law (or 1/$f$-like) scaling of the low-frequency activity was performed (Fig.5 right column). The average Hurst exponent of the temporal dynamics of the RSNs lies in the range of 0.85$-$0.9 (Fig.6\textbf{\textbf{A}}), indicating that the dFCs in infants at term age have persistent, long-range temporal dependencies, which is far from a random process, consistent with slow power-law decay.
\subsubsection*{Temporal dynamics of RSNs reflect fast-slow sequence of cortical network states in infants}
\begin{figure*}[htbp]
{\includegraphics[height=11cm,width=14cm]{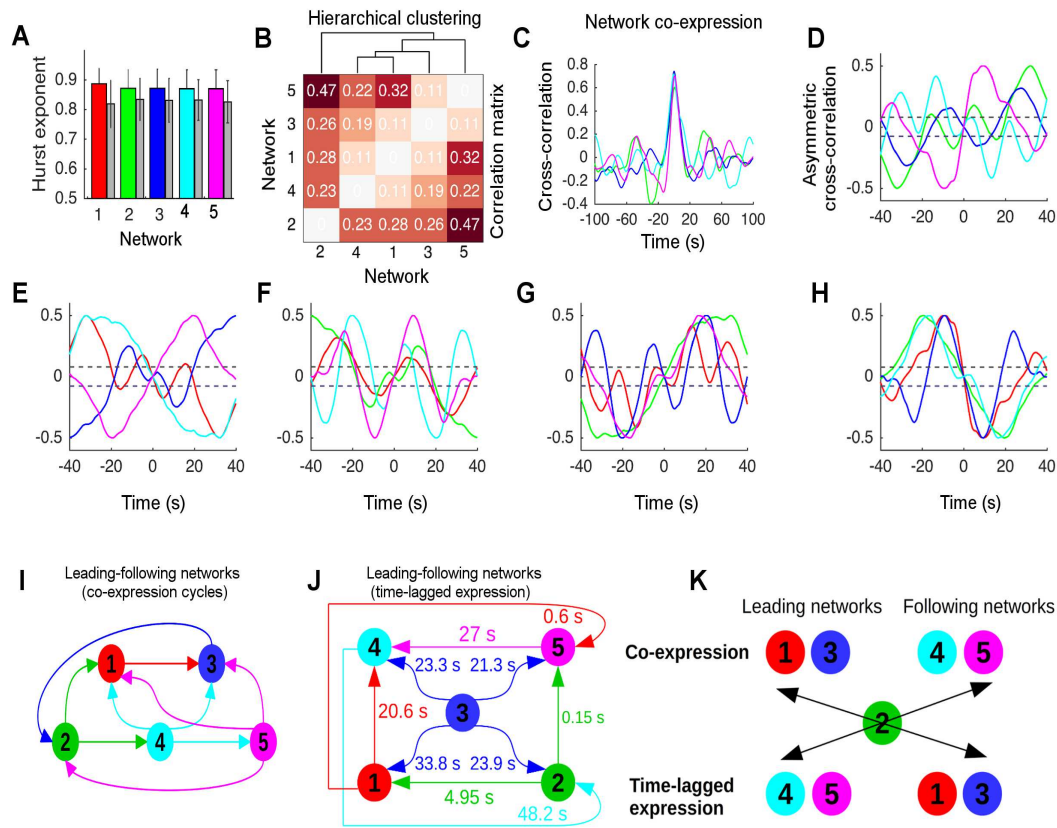}}
\caption{\textbf{Temporal organisation of brain network dynamics in infants.} \textbf{A}, the Hurst exponent of each of the five RSNs. \textbf{B}, hierarchical clustering of the correlation matrix of the networks' temporal-dynamics. \textbf{C}, representative temporally fast and slow network-expressions at $t=0 s$ and $t\neq0 s$ inferred from the cross-correlation between the temporal dynamics of the DMN and the others. \textbf{D-H}, the leading and following properties of the five RSNs were inferred from the analysis of positive and negative slopes of the asymmetric component of the cross-correlations between the networks' temporal dynamics in the time intervals of [-40, 40 s]. The 95\% confidence intervals (dashed-line) for trivial non-zero values due to finite sample length were estimated using a non-parametric permutation test. \textbf{I-J}, respectively show the temporal organisations of RSN interactions in the fast (leading networks 1 and 3, and following networks 2, 4 and 5), and slow sequence of cycles (leading networks 4 and 5 and following networks 1, 2 and 3). \textbf{K}, shows a switch (or a swap) in the dynamics of leading and following networks among the temporally fast- and slow- network expressions with network 2 being as a pass-through state.
}
\label{F6}
\end{figure*}
\noindent The correlation matrix of the RSNs' temporal-dynamics show four levels of hierarchically interconnected clusters (1,000 permutation tests, one-sided $P<0.01$, z-score=2.4) (Fig.6\textbf{\textbf{B}}). The temporal dynamics of the DMN and sensorimotor-thalamic system network form a cluster in level 1, i.e., C1:=(1,4), followed by the three hierarchical levels C2:=(thalamo-frontal network, C1), C3:=(visual-limbic system network, C2) and C4=(primary sensory$-$limbic system network, C3) (Fig.6\textbf{\textbf{B}}). The zero time-lag and non-zero time-lag cross-correlations between the temporal fluctuations of the networks respectively show temporally fast and slow sequences of network expressions in RS cortical activity. That is, dynamics in some networks are expressed slightly earlier than others at around $t=0 s$, while some are expressed with longer time delays than others (e.g. $t>20 s$) (Fig.6\textbf{\textbf{C}}), showing the existence of an asymmetry in the dynamics of interactions among RSNs (panels \textbf{D-H}). As an example, the temporal fluctuation of the thalamo-sensorimotor network is strongly correlated with the others at zero time-lag correlation, slowly decaying toward zero over time delay (Fig.6\textbf{\textbf{C}}) $-$ similar findings pertain to all other networks. Note that, the positive/negative slopes of the temporal asymmetric cross-correlation between the networks at around $t=0 s$ determine the leading-following position of the network under analysis. Each of the two temporally fast and slow sequences of leading-following network interactions show two cycles: A \textit{three-network cycle}, which consists of $\{(1,3,2); (2,4,5)\}$ (panel \textbf{I}) and $\{(1,4,2); (2,5,4)\}$ (panel \textbf{J}), and a \textit{four-network cycle}, which includes $\{1,3,2,4\}; \{2,4,5,3\}$ (panel \textbf{I}) and $\{(1,5,4,2)\}$ (panel \textbf{J}). As an example, the thalamo-frontal network follows the networks 1, 2, 4, and 5 at the time delays of 33.8 $s$, 23.9 $s$, 23.3 $s$ and 21.3 $s$ respectively (panel \textbf{J}). The DMN and thalamo-frontal network, with in-degree=3, are the leading networks in the fast-sequence of network-expression (panel \textbf{I}), tending to follow the sensorimotor-thalamic system and visual-limbic system networks in the slow-sequence of network-expression (panel \textbf{J}), revealing the existence of a swap in the leading and following networks among the temporally fast and slow sequences of network expressions. The primary-sensory limbic system network (i.e. network 2) that has equivalent scores of in-degree=2 and out-degree=2 among the fast- and slow- sequences of network expressions plays a \textit{pass-through network} in the dynamical interplay between the other four networks (panel \textbf{K}).
\newpage
\section*{Discussion}
Our analysis of RS-EEG source reconstruction in 100 infants at term age revealed five robust RSNs that capture, for the first time in the human infants' functional brain network studies, the dominant patterns of anatomically constrained spatially, spectrally and temporally resolved dynamic synchronisation between neuronal populations activity. These RSNs are constituted by a narrowband carrier frequency (1-8 Hz) that is nested by much slow and persistent fluctuations ($<$0.1 Hz). The networks have independent topological organisations, distinct spectral fingerprints and hierarchically correlated temporal dynamics with fast and slow sequences of network states, effectively consistent with dynamic repertoire and macroscopic RS cortical activity in the human adult brain \cite{Vidaurre12827}. Our findings of the DMN, primary functional systems and thalamocortical networks are consistent with the immature DMN, dorsal attention network configuration, and primary functional system networks obtained from the rs-fMRI and diffusion-\\MRI data in infants at term age \cite{Doria201020015, Fransson2011Cerebcortex, Ball20147456, VanDenHeuvel20153000, BATALLE2017379, CaoMiao20172731949}. The existence of the thalamus and basal ganglia in the rich-club of hub organisation in these five RSNs suggests that cortical layers IV and III have respectively established mature functional connections with the thalamus and other cortical layers to regulate the evolution of cortico-cortical functional networks and high order cortices that support salience, executive, integrative, and cognitive functions in infancy and childhood \cite{TauGregory2009, CaoMiao20172731949, GilmoreJohnH2018123}. Co-existence of the small-world topology and vulnerability of these five RSNs suggests that the primary dynamical complexity of the brain functions have the necessary, but insufficient, conditions for information processing at term age, consistent with the topological principles of the RSNs identified by diffusion MRI and fMRI studies in infants \cite{Doria201020015, Fransson2011Cerebcortex, Ball20147456, VanDenHeuvel20153000, CaoMiao20172731949}, however such poor network vulnerability gradually decreases during the first two years of life \cite{Gao2011, CaoMiao20172731949}.\\
These networks have distinct spectral fingerprints with narrowband frequency bases: The DMN has a bell-shape peak within the delta-theta bands, the primary sensory, thalamocortical and limbic system networks have peaks in the narrowband delta-theta and low-alpha frequencies. The delta-band spectral fingerprint of the thalamo-frontal and thalamo-sensorimotor networks provides a convergent evidence that a resonant behaviour and low frequency preferences of neuronal population activity in the thalamus regulate cortical oscillations throughout delta-band thalamocortical network, consistent with previous reports \cite{IgorTimofeev2011112457}. Notably, single peaks in the delta and low-alpha bands associated with the visual-limbic system network suggest the presence of emerging visual network, consistent with the fMRI findings \cite{Fransson15531, Doria201020015} and scalp-EEG power spectrum \cite{CornelissenLaura2015}. The temporal dynamics of the five RSNs counterpart those microstates that were obtained from fluctuations in large-scale field power in the adult scalp-EEG data: The Hurst exponents of the networks temporal-dynamics (0.88$-$0.9) are significantly higher than 0.5 (i.e. random fluctuations), consistent with previous studies \cite{Mehrkanoon2014338}. The co-existence of long-range correlation in network dynamics on slow timescale ($<0.1$ Hz), distinctive spectral fingerprints of the networks and richly structured network topologies suggest that the brain's intrinsic functional organisation underpin the establishment of dynamic functional connectome in infancy, consistent with macroscopic cortical network dynamics that evolve on very slow timescale and have long-range persistence in the adult RS brain network dynamics \cite{Linkenkaer-Hansen20011370, Mehrkanoon2014338}. Networks with continuous oscillatory peaks in the delta-theta bands (such as networks 1 and 3) are typically expressed slightly earlier than others during each fast cycle of network expression (i.e. $t=0 s$), whereas networks with discrete oscillatory peaks in the delta, theta and low-alpha bands are typically expressed over slow cycle of network expression $t\approx25 s$ (such as networks 4 and 5). This timing profile of network dynamics suggests an indirect brain state transition from spectrally and temporally ordered network state to those with discrete oscillatory peaks during each wave of network expression, consistent with the view of ``dynamic FC is the direct product of intrinsic brain electrical activity at distinct frequencies in the adult brain" \cite{TagliazucchiEnzo20126339}. The present findings suggest that infant brain network dynamics arise from long-range synchronisation of bandlimited cortical oscillations subject to an interplay between temporally fast and slow sequences of coherent neuronal populations activity. The timing profile of this interplay is not temporally symmetric among the networks, but rather the appearance of some networks such as the default mode and thalamofrontal characteristically leads the primary-functional$-$limbic, thalamo-sensorimotor and visual-limbic system networks (Fig.6{\textbf{\textbf{I}}}). The temporally fast and slow sequences of network interactions identify the leading and following networks in the evolution of dynamic repertoire of RS cortical activity, by which the temporal formation and dissolution of the brain states are captured (Fig.6 {\textbf{B, I-K}}), reflecting immature meta-state dynamics and evolutionary travelling waves in the infant brain, consistent with the hierarchically organised cortical network dynamics that exhibit spontaneous travelling waves in the adult brain \cite{Vidaurre12827}. Further, spectral fingerprints of the networks identify the neuronal oscillatory mechanisms underpinning the evolution of functional networks. Such functional network evolution is identified by thalamocortical FC (which is part of the networks 1, 3, and 4), and cortico-limbic system (networks 2 and 5) reveal the onset of immature cognitive and memory network formation in infants, which has been shown to have a potential to predict neurocognitive outcomes in children born preterm \cite{Ball20154310}. Furthermore, the timing profile of continuous and discrete oscillatory peaks in the spectral fingerprint of these networks not only offer a system-neuroscience approach to evaluate the neurotypical development and functional brain networks in sick preterm infants, but also provide a capacity for early diagnosis of neurodevelopmental outcomes in disturbed functional brain networks in sick infants.\\
In conclusion, the integration of multilinear PCA and spatial ICA algorithms in conjunction with time-frequency EEG source estimates provides a novel method for inferring patterns of robust dynamic functional brain connectivity in a large cohort of human infants at term-age. These richly organised dynamic brain networks captured the evolution of neuronal populations synchronisation at multiple frequencies. The characteristic temporal dynamics of the networks proved the existence of scale-free property in the neonatal cortical activity in which the brain network states mutually correlate with each other in order to establish functional organisation of the brain in infants. Our findings suggest that the functional organisation of nested resting-state brain network dynamics does not solely depend on the maturation of cognitive networks, instead, the brain network dynamics exist in infants at term-age well before the mature brain networks in the childhood and adulthood periods, and hence the existence of both spontaneous cortical oscillations and network dynamics in infants at term-age offer a quantitative measurement of neurotypical development in infants. Further, our work offers a novel chapter in the assessment of brain network identification and assessments in infants by quantifying anatomically constrained cortical networks, which is the basis vehicle of our future work to help design an effective clinical diagnostic tool for neuro-developmental disability in sick preterm infants.

\section*{Acknowledgments}
This research was genuinely supported by the author's Fellowship Grant AQRF06016-17RD2 funded by the State Queensland Government, Australia. As part of the Financial Incentive Agreement between the University of Queensland and the Queensland Government of Australia, this study uses the previously published EEG data by Professor Paul Colditz$-$ Perinatal Research Centre, Royal Brisbane and Women’s Hospital, Metro North Hospital and Health Service, Brisbane, Australia. The author thanks Professor Paul Colditz for providing EEG data as required by the Fellowship Grant AQRF06016-17RD2.
\section*{Author contribution}
The author designed the research problem, performed research, analysed data, interpreted the results and wrote the paper.

\end{document}